\documentclass[aps,floatfix,amsmath,nofootinbib,amssymb,towcolumn,superscriptaddress]{revtex4}

\usepackage{overpic}
\usepackage{amssymb}
\usepackage{indentfirst}
\usepackage{feynmf}   
\usepackage{slashed}  
\usepackage{cases}
\usepackage{color}
\usepackage{multirow}
\usepackage{epstopdf}
\usepackage{graphicx,color,bm}
\usepackage{epstopdf}
\usepackage{booktabs}

\usepackage[colorlinks,
            citecolor=green,
            anchorcolor=red,
            menucolor=red,
            linkcolor=red,
            filecolor=red,
            runcolor=red,
            urlcolor=blue,
            frenchlinks=red]{hyperref}

\begin{document}

\title{T-even transverse momentum dependent gluon fragmentation functions in a spectator model}

\author{Xiupeng Xie}\affiliation{School of Physics, Southeast University, Nanjing
211189, China}

\author{Zhun Lu}
\email{zhunlu@seu.edu.cn}
\affiliation{School of Physics, Southeast University, Nanjing 211189, China}

\begin{abstract}
We present a model calculation of transverse momentum dependent (TMD) gluon fragmentation functions for the spin-1/2 and spin-0 hadrons. The model is based on the assumption that a time-like off-shell gluon can fragments into a hadron and a single spectator particle. So far such spectator models have only been used to calculate the TMD distribution functions of quark and gluons. The gluon-hadron-spectator coupling is described by an effective vertex containing two form factors. We obtain the analytic expressions for the four T-even TMD fragmentation functions of the gluon. We also present the numerical results for the $z$-dependence and $k_T$-dependence of the fragmentation functions. Our study shows that the effects of these fragmentation functions may be significant and could be probed by future experimental measurements.

\end{abstract}

\maketitle

\section{Introduction}

Understanding the parton structure of hadrons and the mechanism of hadronization is one of the important tasks in particle physics.
The fundamental theory for studying strong interaction is
Quantum chromodynamics (QCD), in which the gluon plays the role of the gauge boson.
Due to the asymptotic freedom of QCD~\cite{Gross:1973id,Politzer:1973fx}, a number of high-energy scattering processes can be analyzed using perturbation theory.
At high energy scale, hadrons can be viewed as composed of nearly non-interacting partons, and gluon is a dynamical component.
In most cases such analyses are in the form of factorization theorem, which separates the perturbatively calculable part of the cross section from the non-perturbative part~\cite{Collins:1981uw,Collins:1989gx}.
In the case the hadron is in the initial state, the parton distribution functions (PDFs) enter the description of the processes.
If a specific hadron is identified in the final state in inclusive or semi-inclusive processes, parton fragmentation functions (FFs)~\cite{Metz:2016swz} appear frequently as non-perturbative ingredient of the factorization framework.

In fact, the PDFs/FFs are essential ingredients in the description of deep inelastic scattering, Drell-Yan, and inclusive hadron production in hadron-hadron collision.
These objects describe the partonic structure involving hadrons.
For instance, the gluon FFs describe how the color-carrying gluons transforms into color-neutral particles such as hadrons or photon.
Unfortunately, the PDFs and FFs still cannot be calculated from first principle because one cannot calculate the soft parts in perturbative QCD.
At present the soft parts of scattering may be treated via lattice calculations~\cite{Ji:2013dva, Ji:2014gla} or phenomenological models.
Alternatively, PDFs and FFs have been extracted  \cite{Martin:2009iq,Buckley:2014ana,NNPDF:2017mvq,Hou:2019efy,Kretzer:2000yf,deFlorian:2007aj,deFlorian:2007ekg,Albino:2008fy, Bertone:2017tyb} from experimental data by global fitting.

If the transverse momentum is probed, the transverse momentum dependent (TMD) PDFs and FFs can emerge in the description of in high energy process involving hadrons.
Particularly, the unpolarized TMD FF $D_1^{h/a}(z, \bm P_{hT}^2)$ describes the probability density of a parton $a$ fragmenting to a hadron $h$ with longitudinal momentum fraction $z$ and transverse momentum $P_{hT}$ with respect to the parent parton.
Compared to the quark TMD PDFs/FFs~\cite{Bacchetta:2006tn,Metz:2016swz} and gluon TMD PDFs, the information of the gluon TMD FFs~\cite{Mulders:2000sh} still remain less know.
On the one hand, the experimental data that constrain the gluon TMD FFs is very limited.
on the other hand, from the theoretical point of view, it is not so clear how to generate the hadronic degree of freedom from a colored gluon.

In this work, we study the gluon TMD FFs using a phenomenological approach.
In order to generate hadronic degree of freedom, we apply the spectator model in the study.
This model, particularly for calculating the gluon TMD FFs, is based on the assumption that a (time-like off-shell) gluon can fragment into a hadron and a single (real) spectator particle.
It allows us to compute TMD FFs in a intuitive way.
The spectator model in different forms was widely applied to calculate the quark TMD PDFs of the nucleon~\cite{Jakob:1997wg,Brodsky:2002cx,Gamberg:2003ey,Bacchetta:2003rz,Bacchetta:2008af}, the quark TMD PDFs of the pion~\cite{Lu:2004hu,Meissner:2008ay,Ma:2019agv}, and the TMD FFs of pion and kaon~\cite{Bacchetta:2007wc}.
Recently, the spectator model has been also discussed in the calculation of the gluon TMD PDFs~\cite{Lu:2016vqu,Bacchetta:2020vty}.
Here, we extend the model to calculate the
leading-twist time-reversal-even (T-even) FFs: $D_{1}^{h / g}$, $G_{1 L}^{h / g}$, $G_{1 T}^{h / g}$, and $H_{1}^{\perp h / g}$.
The first is unpolarized, while the other three FFs are polarized FFs.
Furthermore, we consider the cases the final state hadron is a proton and a pion.
The gluon-hadron-spectator coupling is described by an effective vertex containing two form factors.
The parameters of the model are fixed by fitting the model result for the unpolarized FF $D_1^{h/g}(z)$ to the available parametrization.

The rest of the paper is organized as follows. In Section. II, we provide the formalism of the spectator model for the gluon TMD FFs.
We obtain the model result of the gluon-gluon correlator at tree level and get the leading-twist T-even gluon TMD FFs using proper projecting operators.
In Section. III, we perform a fit of the unpolarized FF to determine the values of the parameters. 
We then provide the numerical prediction for the FFs $G_{1L}^{p/g}$, $G_{1T}^{p/g}$,  $H_1^{\perp, p/g}$ and $H_1^{\perp, \pi/g}$.
We summarize the paper in Section. IV.

\section{Analytic calculation of the T-even FFs of spin-1/2 and spin-0 hadrons}

Leading and non-leading contributions to the hadronic tensor can be identified if one uses a suitable parametrization of the hadron momentum and spin vectors in terms of two light-like directions, i.e., light-cone decomposition. 
Generally, a light-cone vector $a_\mu$ can be expressed as $[a_-, a_+, \bm{a_T}]$ or $a^{-}n_{-}+a^{+}n_{+}+\bm{a_T}$, where $n_-$ and $n_+$ are two light-cone vectors (such that $n_+ \cdot n_- =1$ and $n_\pm^2=0$), and $\bm{a_T}$ is a transverse component. 
In a reference frame in which the hadron has no transverse momentum, one can write:
\begin{align}
P_{h} & = P_{h}^{-} n_{-}+\frac{M_{h}^{2}}{2 P_{h}^{-}} n_{+}\,,\\
k & = \frac{P_{h}^{-}}{z} n_{-}+\frac{z\left(k^{2}+\boldsymbol{k}_{T}^{2}\right)}{2 P_{h}^{-}} n_{+}+k_{T}\,,\\
S_{h} & = S_{h L} \frac{P_{h}^{-}}{M_{h}} n_{-}-S_{h L} \frac{M_{h}}{2 P_{h}^{-}} n_{+}+S_{h T}\,,
\end{align}
where $M_{h}$ is the mass of the final-state hadron and evidently $z=P_{h}^{-} / k^{-}$ is the momentum fraction carried by the hadron, $\bm k_T$ is the transverse momentum of the gluon with respect to $\bm P_h $.
In the case $h$ is polarized such as proton, a gluon with momentum $k$ fragments into a hadron with momentum $P_h$ and spin $S_h=1/2$.

The appropriate gauge-invariant gluon-gluon correlator for fragmentation can be expressed as~\cite{Mulders:2000sh,Metz:2016swz}
\begin{align}
\Delta^{\mu \nu ; \rho \sigma}\left(k ; P_{h}, S_{h}\right) & = \sum_{X} \int \frac{d^{4} \xi}{(2 \pi)^{4}} e^{i k \cdot \xi}\left\langle 0\left|F^{\rho \sigma}(\xi)\right| P_{h}, S_{h} ; X\right\rangle\left\langle P_{h}, S_{h} ; X\left|\mathcal{U}(\xi, 0) F^{\mu \nu}(0)\right| 0\right\rangle\,.
\end{align}
In this work, we use the notation $F_{\mu \nu}(\xi) \equiv F_{\mu \nu}^{a}(\xi) T^{a}$, related to the potential by $F_{\mu \nu}=\partial_{\mu} A_{\nu}-\partial_{\nu} A_{\mu}-i g\left[A_{\mu}, A_{\nu}\right]$. $\mathcal{U}(\xi, 0)$ is the gauge-link operator connecting space-times $\xi$ and 0 to ensure the gauge-invariance of the operator definition. 
For our purpose we calculate the T-even fragmentation function, only the leading-order contributions to the correlator are considered. 
Therefore the effects of the gauge-link are neglected in the present work.

For the description of fragmentation in leading order in the inverse hard scale $Q$, we need to integrate over the momentum in the light-cone direction for correlation function, with the choice for $P_{h}$, being the momentum component $k^+$,
\begin{align}
\Delta^{h / g, i j}\left(z, \boldsymbol{k}_{T}\right) & = \int d k^{+} \Delta^{-j ;-i}\left(k ; P_{h}, S_{h}\right)\,,
\end{align}
where $i$ and $j$ are transverse spatial indices. 
In leading-twist the above correlator can be decomposed further as
\begin{align}
\delta_{T}^{i j} \Delta^{h / g, i j}\left(z, \vec{k}_{T} ; P_{h}, S_{h}\right)=& 2 P_{h}^{-}\left[D_{1}^{h / g}\left(z, z^{2} \vec{k}_{T}^{2}\right)+\frac{\varepsilon_{T}^{i j} k_{T}^{i} S_{h T}^{j}}{M_{h}} D_{1 T}^{\perp h / g}\left(z, z^{2} \vec{k}_{T}^{2}\right)\right]\,,\label{eq:dD} \\
i \varepsilon_{T}^{i j} \Delta^{h / g, i j}\left(z, \vec{k}_{T} ; P_{h}, S_{h}\right)=& 2 P_{h}^{-}\left[\Lambda_{h} G_{1 L}^{h / g}\left(z, z^{2} \vec{k}_{T}^{2}\right)+\frac{\vec{k}_{T} \cdot \vec{S}_{h T}}{M_{h}} G_{1 T}^{h / g}\left(z, z^{2} \vec{k}_{T}^{2}\right)\right]\,, \label{eq:dG} \\
\hat{S} \Delta^{h / g, i j}\left(z, \vec{k}_{T} ; P_{h}, S_{h}\right)=& 2 P_{h}^{-} \hat{S}\left[\frac{k_{T}^{i} \varepsilon_{T}^{j k} S_{h T}^{k}}{2 M_{h}} H_{1 T}^{h / g}\left(z, z^{2} \vec{k}_{T}^{2}\right)+\frac{k_{T}^{i} k_{T}^{j}}{2 M_{h}^{2}} H_{1}^{\perp h / g}\left(z, z^{2} \vec{k}_{T}^{2}\right)\right. \notag \\
&\left.+\frac{k_{T}^{i} \varepsilon_{T}^{j k} k_{T}^{k}}{2 M_{h}^{2}}\left(\Lambda_{h} H_{1 L}^{\perp h / g}\left(z, z^{2} \vec{k}_{T}^{2}\right)+\frac{\vec{k}_{T} \cdot \vec{S}_{h T}}{M_{h}} H_{1 T}^{\perp h / g}\left(z, z^{2} \vec{k}_{T}^{2}\right)\right)\right]\,.\label{eq:dH}
\end{align}
Here, $\delta_{T}^{i j}$ and $\epsilon_{T}^{i j}$ are the symmetric and anti-symmetric transverse tensors, respectively; and $\hat{S}$ is a symmetrization operator for a generic tensor $O^{i j}$. 
They are defined as:
\begin{align}
-\delta_{T}^{i j} =  g_{T}^{i j} \equiv & g^{i j}-n_{+}^{i} n_{-}^{j}-n_{+}^{j} n_{-}^{i}\,,\\
\epsilon_{T}^{i j} \equiv & \epsilon^{n_{+} n_{-} i j}=\epsilon^{-+i j}\,,\\
\hat{S} O^{i j} \equiv & \frac{1}{2}\left(O^{i j}+O^{j i}-\delta_{T}^{i j} O^{k k}\right)\,.
\end{align}

There are totally eight TMD FFs in Eqs~(\ref{eq:dD}-\ref{eq:dH}).
Among them, $D_{1}^{h / g}$ is the unpolarized gluon fragmentation function, $G_{1 L}^{h / g}$ denotes the longitudinally polarized fragmentation function, $G_{1 T}^{h / g}$ represents the longi-transverse fragmentation function, and $H_{1}^{\perp h / g}$ is linearly polarized fragmentation function, respectively.
These four are T-even.
The other four functions $ D_{1 T}^{\perp h / g}$, $H_{1 T}^{h / g}$, $H_{1 L}^{\perp h / g}$, $H_{1 T}^{\perp h / g}$  are naively T-odd gluon FFs and are not calculated in this work.

In the spectator model for the T-even FFs, the correlator can be modeled as :
\begin{align}
\Delta^{i j}\left(z, \boldsymbol{k}_{T}; S_h\right) \sim &\frac{1}{(2 \pi)^{3}} \frac{1}{2(1-z) k^{-}}\left[\overline{\mathcal{M}}_0^{j}(z, \boldsymbol{k}_{T};S_h) \mathcal{M}_0^{i}(z, \boldsymbol{k}_{T};S_h)\right]\,,
\label{eq:proGamma}
\end{align}
where $\mathcal{M}_0^{i}(k, P_h; S_h)$ is the leading-order amplitude of $g\to h+X$, that is, the transition of the gluon to the hadron and the spectator. 
It has the form
\begin{align}
\mathcal{M}^{i}(z, \boldsymbol{k}_{T}; S_h) &=\left\langle k-P_{h}\left|F_{a}^{-i}\right| P_{h}, S_h\right\rangle \notag \\
&=\bar{u}_{c}(k-P_{h}) \frac{G_{a b}^{i \mu}(k, k)}{k^{2}} \mathcal{Y}_{\mu, b c} U(P_{h}, S_{h})\,,
\label{eq:M}
\end{align}
where $U$ is the spinor for the spin-1/2 hadron and $\bar{u}_{c}$ is the spinor of the spectator with color $c$. 
The term
\begin{align}
G_{a b}^{i \mu}(k, k) & = -i \delta_{a b}k^{-}\left(g^{i \mu}-\frac{k^{i}n_{+}^{\mu}}{k^{-}}\right)
\end{align}
corresponds to the Feynman rule for the field strength tensor of the form $-i\left(p^{\mu} g^{\nu \rho}-p^{\nu} g^{\mu \rho}\right) \delta_{a b}$~\cite{Goeke:2006ef,Collins:2011zzd}, and $\mathcal{Y}_{b c}^{\mu}$ is the gluon-hadron-spectator vertex.

Using Eq.~(\ref{eq:proGamma}) and Eq.~(\ref{eq:M}), we can write our spectator model approximation to the gluon-gluon correlator at tree level as
\begin{align}
\Delta^{i j}\left(z, \boldsymbol{k}_{T}, S_h\right)=& \frac{1}{(2 \pi)^{3}} \frac{1}{2(1-z) k^{-}} \operatorname{Tr}\left[(\not{P_h}+M_h) \frac{1+\gamma^{5}\not{S_h}}{2}\right.\left. G_{a b^{\prime}}^{j v *}(k, k) \mathcal{Y}_{v, b^{\prime} c^{\prime}}^{*} G_{a b}^{i \mu}(k, k) \mathcal{Y}_{\mu, b c}\left(\not{k}-\not{P_h}+M_{X}\right)_{c c^{\prime}}\right]\,,
\label{eq:proGamma2}
\end{align}
where a trace over the Dirac space for spin-1/2 hadron is understood.

For the gluon-hadron-spectator vertex, similar to the choice in Ref.~\cite{Bacchetta:2020vty}, we adopt the following form
\begin{align}
\mathcal{Y}_{b c}^{\mu} & = \delta_{b c}\left[g_{1}\left(k^{2}\right) \gamma^{\mu}+g_{2}\left(k^{2}\right) \frac{i}{2 M_{h}} \sigma^{\mu \nu} k_{\nu}\right]\,,
\end{align}
where $\sigma^{\mu \nu}=i\left[\gamma^{\mu}, \gamma^{\nu}\right]/2$, $g_{1}(k^{2})$ and $g_{2}(k^{2})$ are the gluon-hadron-spectator couplings.
There are several different choices in the literature~\cite{Bacchetta:2008af} for the kind of coupling.
The simplest choice is the point-like coupling, i.e, $g_i$ represents a coupling constant.
However, this will lead to divergence when integrating over the transverse momentum $k_T$.
Usually a cut-off $\Lambda=k_T^{\textrm{max}}$ is introduced in this case to regularize the divergence.
On the other hand, the cut-off can be implemented smoothly by choosing the exponential form factor or the dipolar form factors.
In this work, we apply the dipolar form factor following Ref.~\cite{Bacchetta:2020vty},
\begin{align}
g_{1,2}\left(k^{2}\right) & = \kappa_{1,2} \frac{k^{2}}{\left|k^{2}-\Lambda_{X}^{2}\right|^{2}}\,,
\end{align}
where $\kappa_{1,2}$ are free parameters and $\Lambda_{X}$ is cut-off parameter.

Since the spectator is on-shell $(k-P_h)^{2}=M_X^{2}$, one can obtain the following expression for $k^2$:
\begin{align}
k^{2} & = \frac{z}{1-z}\vec{k}_{T}^{2}+\frac{M_X^{2}}{1-z}+\frac{M_{h}^{2}}{z}\,,
\label{eq:k}
\end{align}
where $M_X$ is the mass of the spectator.

By projecting $\Delta^{i j}$~\cite{Mulders:2000sh,Metz:2016swz} with $\delta_{T}^{i j}$, $\varepsilon_{T}^{i j} $ and $\hat{S}$, we obtain the expressions of the leading-twist T-even gluon TMD FFs:
\begin{align}
D_{1}^{h/g}\left(z, \boldsymbol{k}_{T}^{2}\right) =& \frac{1}{2} \delta_{T}^{i j} {\left[\Delta^{i j}\left(z, \boldsymbol{k}_{T}, S_h\right)+\Delta^{i j}\left(z, \boldsymbol{k}_{T},-S_h\right)\right]} \notag \\
=&[z^2 \boldsymbol{k}_T^2 \left(4 g_{1}^2 M_h^2 \left(2 z^2-2 z+1\right)+4 g_1 g_2 M_h (M_X-M_h)+g_2^2 \left(M_h^2 (3-2 z)-2 M_h M_X \right. \right. \notag \\
&\left. \left. +M_X^2 (2 z+1)\right)\right)+(M_h (z-1)+M_X z)^2 (2 g_1 M_h+g_2 (M_X-M_h))^2+2 g_2^2 z^4 \boldsymbol{k}_T^4] \notag \\
&\times \left[64 \pi ^3 M_h^2 \left(z^2 \boldsymbol{k}_T^2+(1-z)M_h^2+zM_X^2 \right)^2\right]^{-1} \,, \label{eq:d1}\\
G_{1 L}^{h/g}\left(z, \boldsymbol{k}_{T}^{2}\right)  =& \frac{1}{S_{hL}} i \varepsilon_{T}^{i j} \Delta^{i j}\left(z, \boldsymbol{k}_{T}, S_{hL}\right) \notag \\
=&-\left[(2 g_1 M_h+g_2 (M_X-M_h)) \left(z^2 \boldsymbol{k}_T^2 (2 g_1 M_h (2 z-1)+g_2 M_h (2 z-3)+g_2 M_X (2 z-1)) \right. \right. \notag\\
&\left. \left. +(M_h (z-1)+M_X z)^2 (2 g_1 M_h+g_2 (M_X-M_h))\right)\right] \notag \\
& \times \left[64 \pi ^3 M_h^2 \left(z^2 \boldsymbol{k}_T^2+(1-z)M_h^2+zM_X^2 \right)^2\right]^{-1}\,,\label{eq:dg1L}\\
G_{1 T}^{h/g}\left(z, \boldsymbol{k}_{T}^{2}\right)  =& \frac{M}{\boldsymbol{k}_{T} \cdot S_{hT}} i \varepsilon_{T}^{i j} \Gamma^{i j}\left(z, \boldsymbol{k}_{T}, S_{hT}\right) \notag \\
=&-\frac{(2 g_1 M_h+g_2(M_X-M_h))\left(z(M_h(z-1)+M_X z)(2 g_1 M_h(z-1)-g_2 M_X)-g_2 z^{3} \boldsymbol{k}_{T}^{2}\right)}{32 \pi^{3} M_h\left(z^{2} \boldsymbol{k}_{T}^{2}+(1-z)M_h^{2}+zM_X^{2} \right)^{2}}\,, \label{eq:dg1T}\\
H_{1}^{\perp h/g}\left(z, \boldsymbol{k}_{T}^{2}\right)  =&\frac{\hat{S} \Delta^{i j}\left(z, \boldsymbol{k}_{T} ; S_{h}\right)}{\hat{S} \frac{k_{T}^{i} k_{T}^{j}}{2 M_{h}^2}} \notag \\
=&\frac{z^2 \left(4 g_1^2 M_h^2 (z-1) z+g_2^2 z^2 \boldsymbol{k}_T^2+g_2^2 \left(M_h^2 (-z)+M_h^2+M_X^2 z\right)\right)}{16 \pi ^3 \left(z^2 \boldsymbol{k}_T^2+(1-z)M_h^2 +zM_X^2 \right)^2}\,. \label{eq:dh1}
\end{align}

For completeness, we also present the calculation of the gluon fragmentation function for spin-0 hadron such as pion meson. 
Similar to Eq.~(\ref{eq:proGamma}), the correlator can be written as
\begin{align}
\Gamma^{i j}\left(z, \boldsymbol{k}_{T}\right) \sim &\frac{1}{(2 \pi)^{3}} \frac{1}{2(1-z) k^{-}}\overline{\mathcal{M}}_{0}^{j}(k, P_h) \mathcal{M}_{0}^{i}(k, P_h) \notag \\
=& \frac{1}{(2 \pi)^{3}} \frac{1}{2(1-z) k^{-}} G_{a b^{\prime}}^{j \nu *}(k, k) \mathcal{Y}_{\nu, b^{\prime} c^{\prime}}^{*} G_{a b}^{i \mu}(k, k) \mathcal{Y}_{\mu, b c}\,,
\end{align}
where the vertex for spin-0 hadron has the form:
\begin{align}
\mathcal{Y}_{b c}^{\mu} & = \delta_{b c} g(k^{2}) \left[P_h^\mu - \left(k^\mu - P_h^\mu\right)\right]\,,
\end{align}
where $g(k^{2})$ the form factor for the gluon-hadron-spectator coupling.

In leading-twist there are two gluon fragmentation functions for spin-0 hadron, $D_{1}^{\pi/g}$ and $H_{1}^{\perp \pi/g}$, their expressions can be obtained by using projecting operators $\delta_{T}^{i j}$ and $\hat{S} $:
\begin{align}
D_{1}^{h/g}\left(z, \boldsymbol{k}_{T}^{2}\right) =&  \delta_{T}^{i j} \Delta_\pi^{i j}\left(z, \boldsymbol{k}_{T}\right) \notag \\
=&\frac{g^2 (1-z) z^3 \boldsymbol{k}_T^2}{8 \pi ^3 \left(z^2 \boldsymbol{k}_T^2+M_pi^2 (1-z)+M_X^2 z\right)^2}\,, \label{eq:pion1}\\
H_{1}^{\perp \pi/g}\left(z, \boldsymbol{k}_{T}^{2}\right)  =&\frac{\hat{S} \Delta_\pi^{i j}\left(z, \boldsymbol{k}_{T} \right)}{\hat{S} \frac{k_{T}^{i} k_{T}^{j}}{2 M_{pi}^2}} \notag \\
=&\frac{g^2 M_\pi^2 (1-z) z^3}{4 \pi ^3 \left(z^2 \boldsymbol{k}_T^2+M_\pi^2 (1-z)+M_X^2 z\right)^2}\,. \label{eq:pion2}
\end{align}
It is interesting to point out that, from (\ref{eq:pion1}) and (\ref{eq:pion2}), one can find a relation between
$H_1^{\perp h/g}(z,\bm k_T^2)$ and $D_1^{h/g}$:
\begin{align}
{\bm k_T^2 \over 2M_\pi^2}H_{1}^{\perp h/g}(z,\bm k_T^2)= D_1^{h/g}(z,\bm k_T^2)\,. \label{eq:pi_relation}
\end{align}
This is of course a model-dependent relation which is only valid in this particular model.

\section{Numerical Results}

\begin{table}[h]
\centering
\renewcommand\tabcolsep{5.0pt}
\renewcommand{\arraystretch}{2}
\begin{tabular}{cc}
\toprule[2pt]
parameter              & value           \\
\midrule[1pt]
$\kappa_{1}^{p}$       & 3.769$\pm$1.306 \\
$\kappa_{2}^{p}$       & 6.785$\pm$1.644 \\
$\Lambda_{X}^{p}$      & 2.224$\pm$0.044 \\
$M_X^{p}$            & 1.589$\pm$0.071 \\
\midrule[1pt]
$\kappa^{\pi}$       & 4.268$\pm$0.708  \\
$\Lambda_{X}^{\pi}$  & 0.477$\pm$0.019  \\
$M_X^{\pi}$        & 0.500$\pm$0.033  \\
\bottomrule[2pt]
\end{tabular}
\caption{first four rows: parameters for the proton; last three rows: parameters for the pion meson.}
\label{table:para}
\end{table}

\begin{figure}
\centering
\includegraphics[width=0.45\columnwidth]{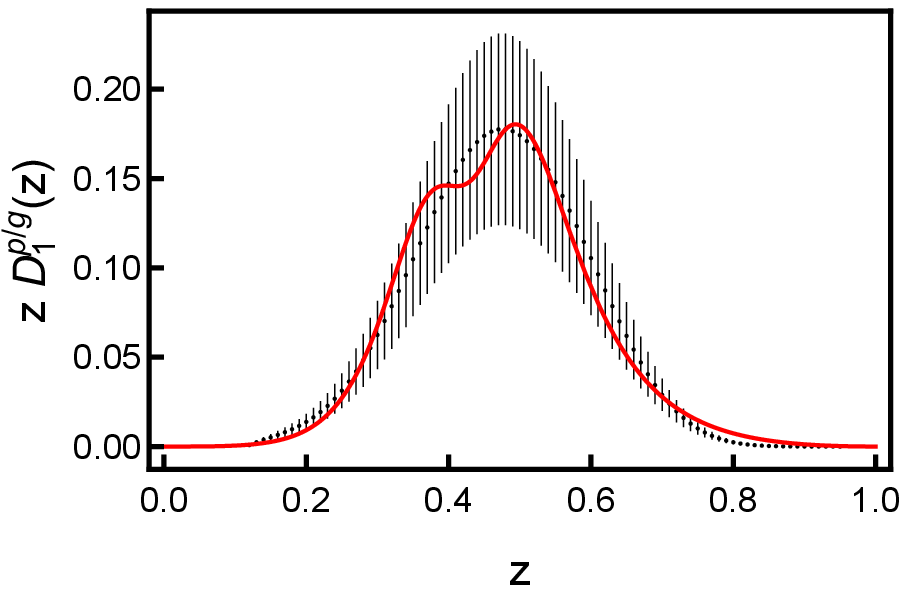}
\quad
\includegraphics[width=0.45\columnwidth]{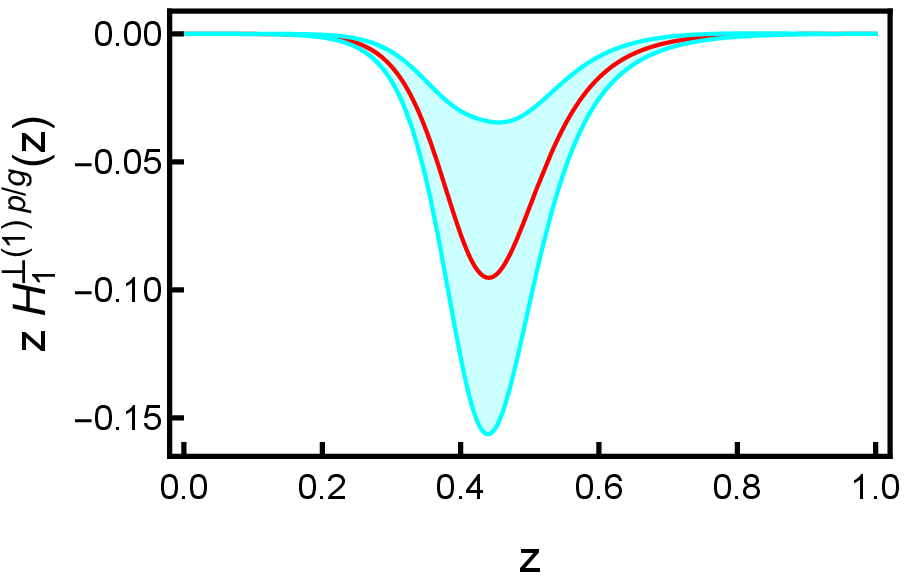}
\quad
\includegraphics[width=0.45\columnwidth]{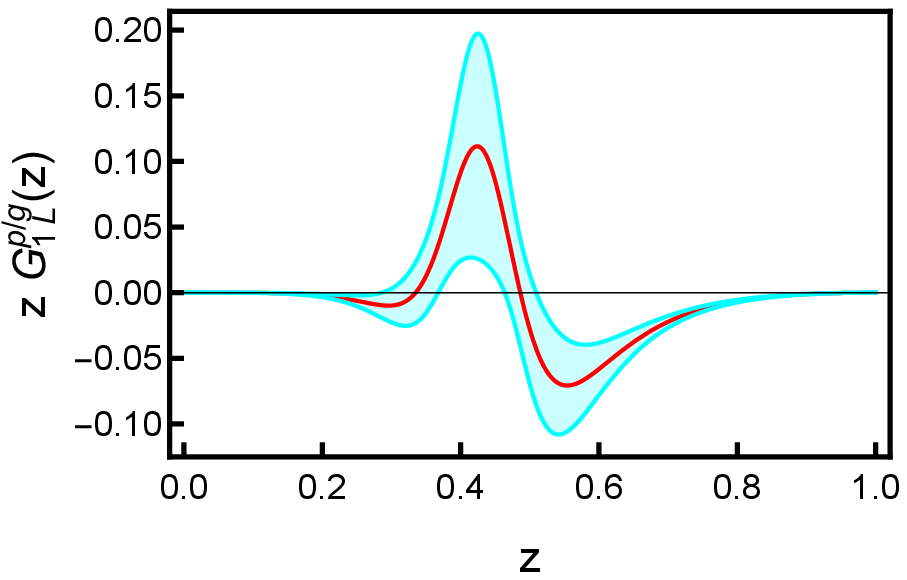}
\quad
\includegraphics[width=0.45\columnwidth]{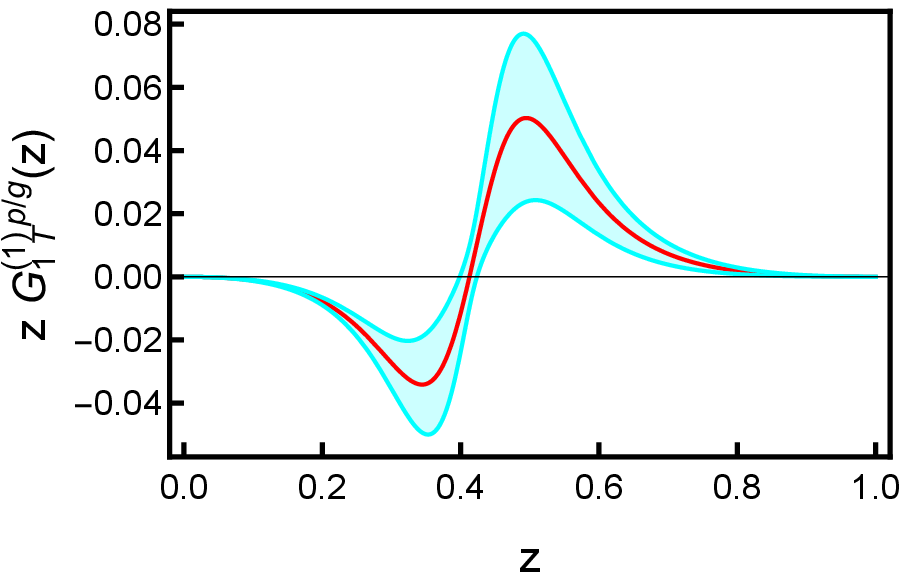}
\caption{Upper-left panel: Fit of the unpolarized gluon fragmentation function for the proton as a function of $z$. 
The solid circles correspond to the AKK08 parametrization, the error bars represent the uncertainties of the fragmentation function (assumed as $20\%$ of the value of the AKK08 fragmentation function). 
The solid line depicts the spectator model result.
Other three panels: the central lines depict the z-dependence of $zG_{1}(z)$, $zG_{1 T}^{(1) p/g}(z)$ and $zH_{1}^{\perp (1) p/g}(z)$ in the spectator model calculated from the fitted parameters. The bands depict the uncertainties from the uncertainties of the parameters.}
\label{Fig:proton}
\end{figure}

\begin{figure}
\centering
\includegraphics[width=0.45\columnwidth]{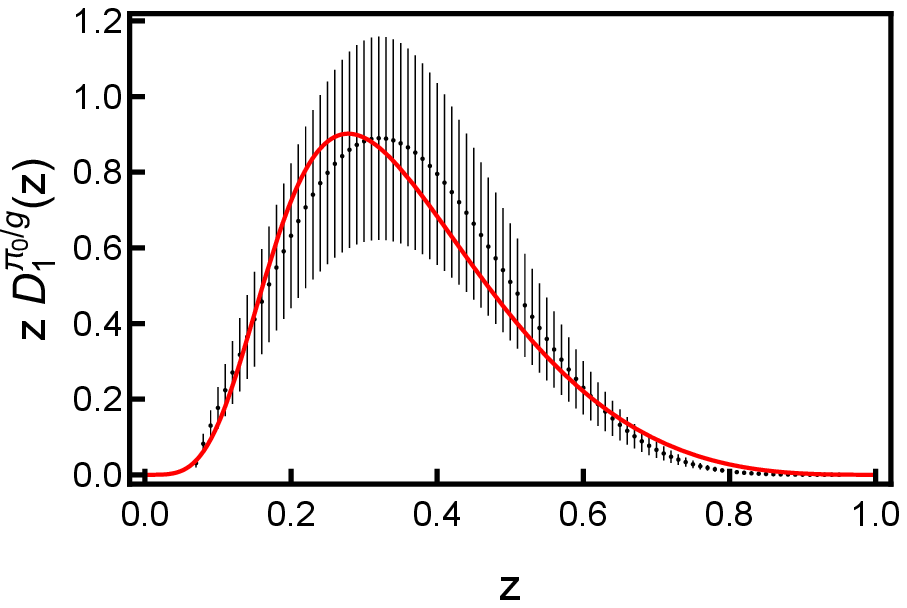}
\quad
\includegraphics[width=0.45\columnwidth]{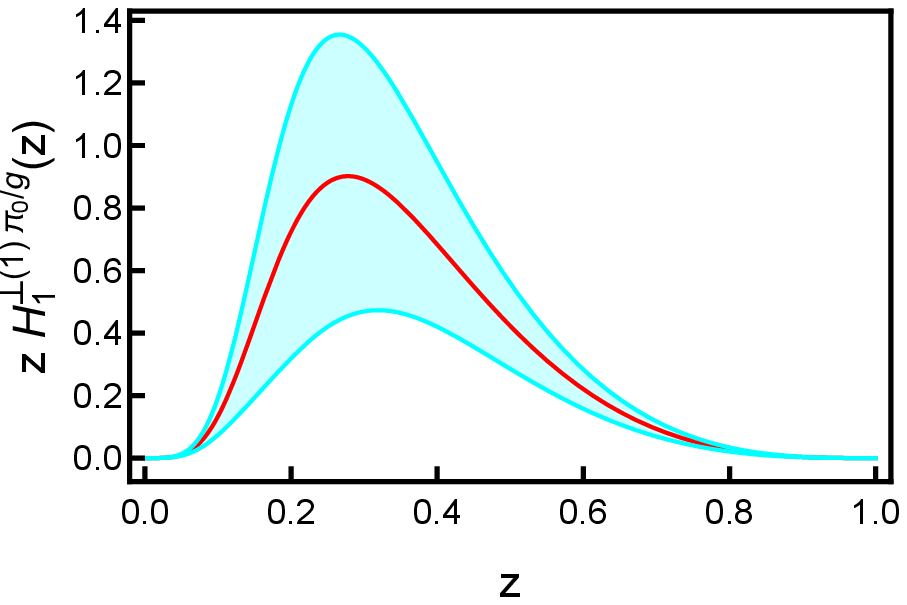}
\caption{Left panel: Fit of the unpolarized gluon fragmentation function for the pion as a function of $z$. The solid circles correspond to the AKK08 parametrization, the error bars represents the uncertainties of the fragmentation function (assumed as $20\%$ of the value of the AKK08 fragmentation function). 
The solid line depicts the spectator model result. 
Right panel: the central lines depict the $z$-dependence of $zH_{1}^{\perp (1) \pi/g}(z)$ in the spectator model calculated from the fitted parameters. 
The bands depict the uncertainties from the uncertainties of the parameters.}
\label{Fig:pion}
\end{figure}

In this section we present the numerical results for the gluon fragmentation function for spin-1/2 and spin-0 hadron.
We need to fix the parameters $M_X$, $\kappa$ and $\Lambda_X$ in the expressions (\ref{eq:d1}-\ref{eq:dh1}), (\ref{eq:pion1}) and (\ref{eq:pion2}).
To do this we perform a fit on the integrated fragmentation function $D_1^{h/g}(z)$ defined as
\begin{align}
D_1^{h/g}(z)=   \int d\bm{P}_{hT}^2 D_1^{h/g}(z, \bm {P}_{hT}^2)=z^2\int d^2 \bm{k}_T D_1^{h/g}(z, \bm {k}_T^2) 
\end{align}
with the available parameterizations.
The prefactor $z^2$ in front of the integral comes from the fact that the transverse momentum of the hadron $h$ with respect to the momentum of the gluon $\bm P_{hT}$ satisfies the relation $\bm P_{hT} =-z \bm k_T$.
As an example, we take proton as the spin-1/2 hadron and pion as the spin-0 hadron to perform the calculation.
We adopt the AKK08 parametrization~\cite{Albino:2008fy} for $zD_1^{p/g}(z)$ and $zD_1^{\pi/g}(z)$ at the sale $Q_0$=1.5 GeV  to perform the fit.
In the fit we consider the parameterizations in the range $0.1<z<0.7$.
The fitted values of the parameters with uncertainties are provided in Table.~\ref{table:para}, in which the first four rows are the parameters for the gluon FFs of the proton, while the last tree rows are those of the pion.
The comparison between the model calculation (using the fitted values) and the AKK parameterizations is shown in the upper-left panel of Fig.~\ref{Fig:proton} (for proton) and the left panel of Fig.~\ref{Fig:pion} (for pion).
Since there is no error in the AKK parameterizations, we assign a constant relative error of $20\%$ to $zD_1^{p/g}(z)$ or $zD_1^{\pi/g}(z)$.
The uncertainties of the parameters are thus estimated from this error.

Using the fitted values of parameters, we provide the prediction of the other three T-even gluon FFs in the spectator model.
First we present the integrated FFs defined as
\begin{align}
G_{1}^{h/g}(z)&= z^2\int d^2 \bm{k}_T G_{1L}^{h/g}(z,  \bm {k}_T^2)\,,\\
G_{1T}^{(1) h/g}(z)&= z^2\int d^2 {\bm{k}_T^2 \over 2 M_h^2}\bm{k}_T G_{1T}^{h/g}(z,  \bm {k}_T^2)\,,\\
H_1^{\perp (1) h/g}(z)& = z^2\int d^2 \bm{k}_T {\bm{k}_T^2 \over 2 M_h^2} H_{1}^{h/g}(z,  \bm {k}_T^2)\,,
\end{align}
where the second line provides the first transverse moment of $G_{1T}^{h/g}(z,  \bm {k}_T^2)$, and the third line provides that of $H_{1}^{h/g}(z, \bm {k}_T^2)$.
In the upper-right, lower-left and lower right panels of Fig.~\ref{Fig:proton}, we plot the $z$-dependence of $zH_1^{\perp (1) p/g}(z)$, $zG_{1}^{p/g}(z)$ and $zG_{1T}^{(1)p/g}(z)$ for the proton, respectively. 
In the right panel of Fig.~\ref{Fig:pion}, we plot the $z$-dependence of $zH_1^{\perp (1) \pi/g}(z)$.
The central lines depict the results from the fitted parameters. 
The bands correspond to the uncertainties of the FFs from the uncertainties of the parameters.
The numerical results show that the sizes of $H_1^{\perp (1) p/g}(z)$, $G_{1}^{p/g}(z)$ are comparable to that of $D_1^{p/g}(z)$.
The size of $G_{1T}^{(1) p/g}(z)$ is several times less than that of $D_1^{p/g}(z)$.
Similarly, the size of $H_1^{\perp (1) \pi/g}(z)$ is comparable to that of $D_1^{\pi/g}(z)$.
This indicates that the effects of these FFs may be significant and could be probed by future experimental measurements.
We also find that $H_1^{\perp (1) p/g}(z)$ is negative and $H_1^{\perp (1) \pi/g}(z)$ is positive in the entire $z$ region, while the signs of $G_{1}^{h/g}(z)$ and $G_{1T}^{(1) p/g}(z)$ flip when $z$ increases from lower region to higher region.   
That is, $G_{1}^{h/g}(z)$ tends to be positive in the region $z<0.5$ and turns to negative when $z$>0.5. Similarly, a node appears at $z = 0.4$ for $G_{1T}^{(1) p/g}(z)$ since the sign of $G_{1T}^{p/g}(z)$ is negative in the smaller $z$ region, while it turns to positive in the larger $z$ region.

\begin{figure}
\centering
\includegraphics[width=0.45\columnwidth]{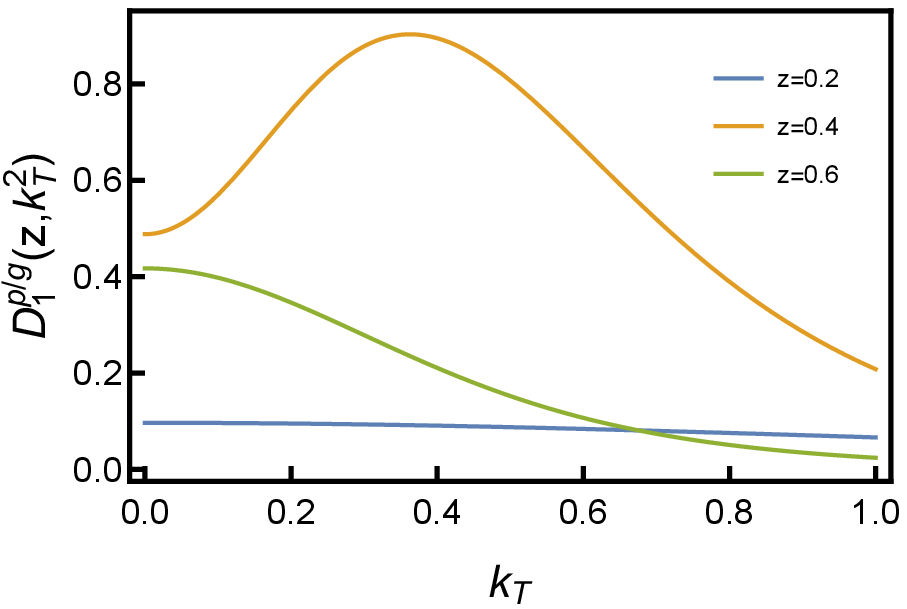}
\quad
\includegraphics[width=0.45\columnwidth]{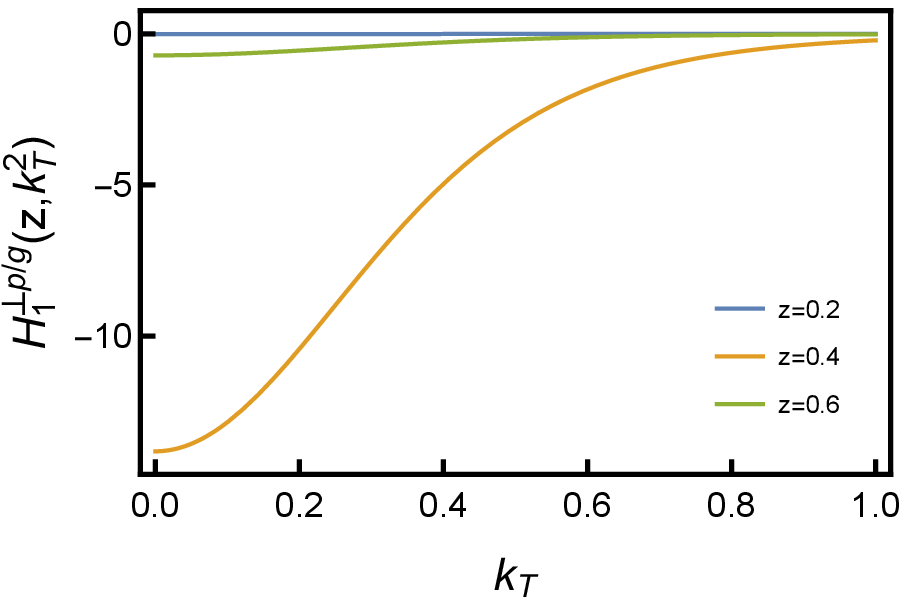}
\quad
\includegraphics[width=0.45\columnwidth]{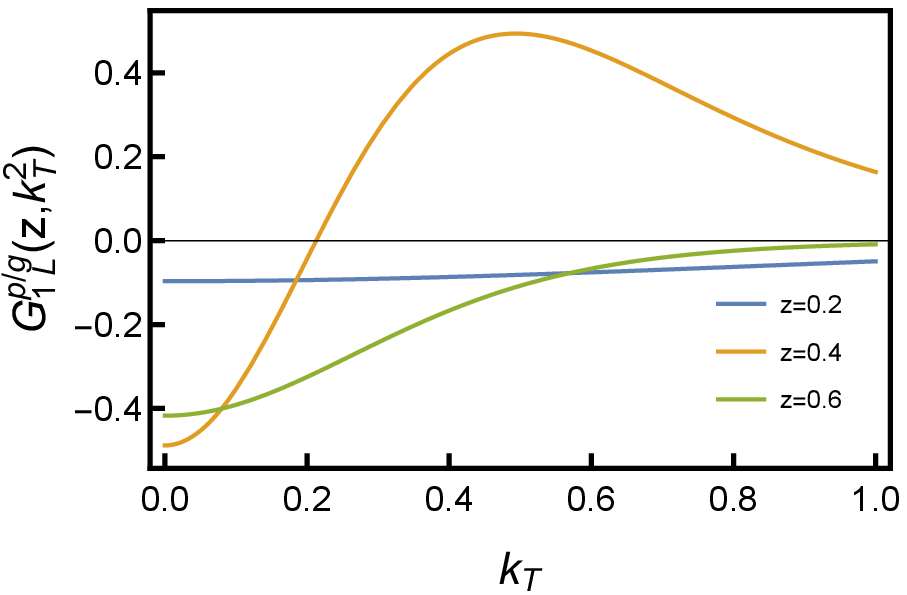}
\quad
\includegraphics[width=0.45\columnwidth]{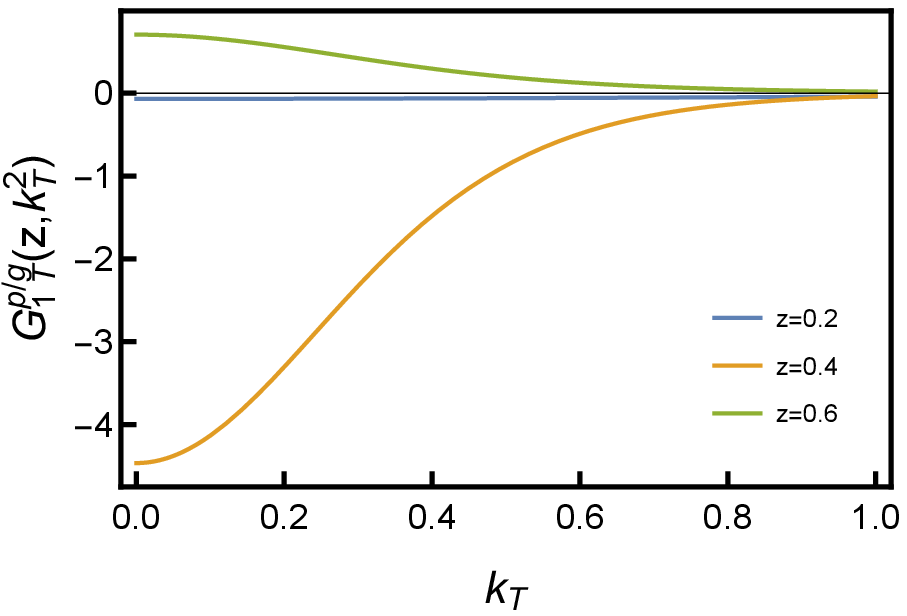}
\caption{ The $k_T$-dependence of the four T-even gluon FFs $D_1^{ p/g}(z,\bm k_T^2)$, $H_1^{\perp (1) p/g}(z,\bm k_T^2)$, $G_{1L}^{p/g}(z,\bm k_T^2)$ and $G_{1T}^{p/g}(z,\bm k_T^2)$ at z=0.2, 0.4 and 0.6, respectively.}
\label{Fig:protonkt}
\end{figure}

\begin{figure}
\centering
\includegraphics[width=0.45\columnwidth]{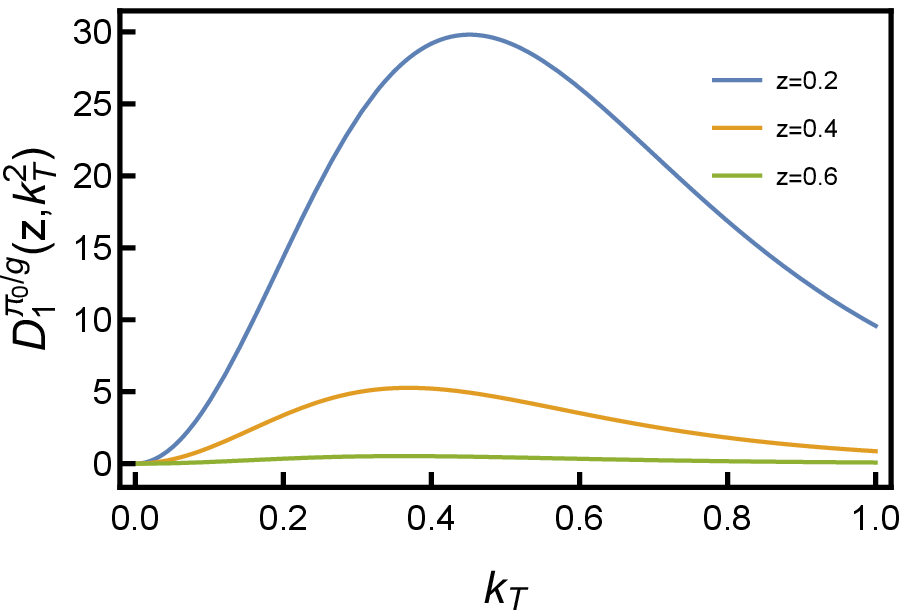}
\quad
\includegraphics[width=0.45\columnwidth]{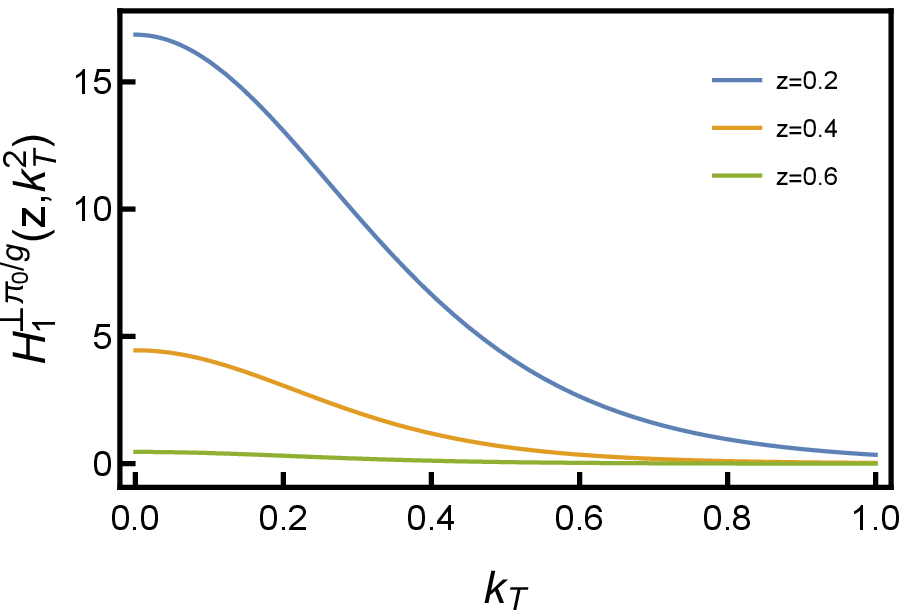}
\caption{The $k_T$-dependence of $D_1^{ pi/g}(z,\bm k_T^2)$ (left) and $H_1^{\perp (1) p/g}(z,\bm k_T^2)$ (right) at z=0.2, 0.4 and 0.6, respectively.}
\label{Fig:pionkt}
\end{figure}

In order to study the $k_T$-dependence of the gluon FFs, we also consider the case the transverse momentum remains unintegrated.
In Fig.~\ref{Fig:protonkt}, we plot the four FFs of the proton: $D_1^{ p/g}(z,\bm k_T^2)$, $H_1^{\perp p/g}(z,\bm k_T^2)$, $G_{1L}^{p/g}(z,\bm k_T^2)$ and $G_{1T}^{p/g}(z,\bm k_T^2)$  as functions of $k_T=|\bm k_T|$ at $z=0.2$, 0.4 and 0.6, respectively.
Similarly, in Fig.~\ref{Fig:pionkt}, we plot the $k_T$-dependence of $D_1^{ \pi/g}(z,\bm k_T^2)$ (left) and $H_1^{\perp (1) \pi/g}(z,\bm k_T^2)$ (right).
We find that the $k_T$-shape of the gluon TMD FFs at different $z$ is different.
For example, at $z=0.2$ and $z=0.6$, $D_1^{ p/g}(z,\bm k_T^2)$ decreases with increasing $k_T$, while at $z=0.4$ there is a peak around $k_T$=0.4 GeV.
$H_1^{\perp p/g}(z,\bm k_T^2)$ is negative in entire $k_T$ region, while $G_{1L}^{p/g}(z,\bm k_T^2)$ can change sign as $k_T$ changes in certain $z$ region.

Positivity bounds~\cite{Bacchetta:1999kz,Mulders:2000sh} provide important model-independent constraints for both the PDFs and FFs in leading-twist.
Particularly, the gluon FFs should satisfy the following bounds~\cite{Mulders:2000sh}
\begin{align}
|G_1^{h/g}(z)|&\leq D_1^{h/g}(z)\,,\label{eq:bound1}\\
{\bm k_T^2 \over 2M_h^2}\left|H_1^{\perp h/g}(z,\bm k_T^2)\right|&\leq D_1^{h/g}(z,\bm k_T^2)\,,\label{eq:bound2}\\
{ |\bm k_T| \over M_h}\left|G_{1T}^{h/g}(z,\bm k_T^2)\right|&\leq D_1^{h/g}(z,\bm k_T^2)\,.\label{eq:bound3}
\end{align}
We have checked numerically that our model results for $G_1^{p/g}(z)$, $H_1^{\perp p/g}(z,\bm k_T^2)$, $G_{1T}^{p/g}(z,\bm k_T^2)$ and $H_1^{\perp \pi/g}(z,\bm k_T^2)$ obeys the bounds in Eqs.~(\ref{eq:bound1}-\ref{eq:bound3}).
An extreme case is the fragmentation function $H_1^{\perp \pi/g}(z,\bm k_T^2)$ in the model, which saturates the positivity bound (\ref{eq:bound2}), as can be seen from the model relation in Eq.~(\ref{eq:pi_relation}).

\section{Conclusions}

We studied the leading-twist T-even TMD gluon FFs in a spectator model.
In the model, the parton is assumed to fragment into a hadron plus a spectator in a single step.
We calculated the gluon-gluon correlators for the fragmentation to a spin-1/2 hadron as well as to a spin-0 hadron.
We obtained the analytic expressions of the four T-even TMD gluon fragmentation function by projecting the correlators to the symmetric and antisymmetric tensor $\delta_T^{ij}$, $\epsilon_T^{ij}$ and $\hat S$.
In the calculation we adopted a dipole form factor for the gluon-hadron-spectator coupling.
Using the AKK08 parametrizations for the unpolarized gluon fragmentation function of the proton and that of the pion, we performed a fit to determine the values of the parameters of the model.
The parameters were applied to predict the $z$-dependence and the $k_T$ dependence of the FFs $G_{1L}^{p/g}$, $G_{1T}^{p/g}$,  $H_1^{\perp, p/g}$ and $H_1^{\perp, \pi/g}$.
We also checked the positivity bounds for the gluon fragmentation function and found that our results satisfy the these bounds.
Our studied showed that the effects of these FFs may be significant and could be probed by future experimental measurements.
Finally, the same model can be also extended to calculate the four leading-twist T-odd gluon TMD FFs.

\end{document}